# Reconstruction of non-classical cavity field states with snapshots of their decoherence

Samuel Deléglise[1], Igor Dotsenko[1,2], Clément Sayrin[1], Julien Bernu[1], Michel Brune[1], Jean-Michel Raimond[1] & Serge Haroche[1,2]

**The state of a microscopic system encodes its complete quantum description, from which the probabilities of all measurement outcomes are inferred. Being a statistical concept, the state cannot be obtained from a single system realization. It can be reconstructed[1] from an ensemble of copies, by performing measurements on different realizations[2-4]. Reconstructing the state of a set of trapped particles shielded from their environment is an important step for the investigation of the quantum to classical boundary[5]. While trapped atom state reconstructions[6-8] have been achieved, it is challenging to perform similar experiments with trapped photons which require cavities storing light for very long times. Here, we report the complete reconstruction and pictorial representation of a variety of radiation states trapped in a cavity in which several photons survive long enough to be repeatedly measured. Information is extracted from the field by atoms crossing the cavity one by one. We exhibit a gallery of pictures featuring coherent states[9], Fock states with a definite photon number and Schrödinger cat states which are superpositions of coherent states with different phases[10]. These states are equivalently represented by their density matrices in the photon-number basis or by their Wigner functions, which are distributions of the field complex amplitude[11]. Quasi-classical coherent states have a Gaussian-shaped Wigner function while Fock and Schrödinger cat Wigner functions show oscillations and negativities revealing quantum interferences. Cavity damping induces decoherence which quickly washes out the Wigner functions oscillations[5]. We observe this process and realize movies of decoherence by reconstructing snapshots of Schrödinger cat states at successive times. Our reconstruction procedure is a useful tool for further decoherence and quantum feedback studies of fields trapped in one or two cavities.**

Engineering and reconstructing non-classical states of trapped light requires cavities preventing the escape of a single photon during the preparation and read-out procedures. We have built a cavity made of highly reflecting superconducting mirrors[12] whose long damping time, $T_c$=0.13 s, allows the trapped field to interact with thousands of atoms crossing it one by one. The interaction with atoms is used to turn an initial coherent field into a Fock or Schrödinger Cat (SC) state and, subsequently, to reconstruct it. An ensemble of trapped photons becomes, like a collection of trapped atoms, an "object of investigation" to be manipulated and observed for fundamental tests and quantum information purposes.

Our set-up is sketched in Fig. 1a. The cavity C, resonant at 51 GHz, is cooled to a temperature of 0.8 K (mean number of residual blackbody photons $n_b$ = 0.05). A coherent microwave field with a Poisson photon number distribution (mean $n_m$, standard deviation $\Delta n$ = $\sqrt{n_m}$) is initially injected in C using a classical pulsed source S. Rubidium atoms from an atomic beam are prepared in box B into the circular Rydberg state with principal quantum number 50 ($|g\rangle$). The cavity is detuned from the transition between $|g\rangle$ and the adjacent circular state 51 ($|e\rangle$) by an amount δ, precluding atom-field energy exchange. The pulsed atom preparation produces Rydberg atoms with a 250 m/s velocity. Auxiliary microwave cavities $R_1$ and $R_2$ sandwiching C are connected to a microwave source S'. They are used to apply resonant pulses to the atoms. The $R_1$ pulse performs the $|g\rangle \rightarrow (|e\rangle + |g\rangle)/\sqrt{2}$ transformation. The same pulse, differing by an adjustable phase-shift ϕ, is applied in $R_2$. Atoms are counted by the detector D discriminating $|e\rangle$ and $|g\rangle$ (one atom on average every 0.5ms). For experimental details, see refs 13 and 14.

The $R_1$-$R_2$ combination forms a Ramsey interferometer[14]. It is sensitive to the atomic state superposition phase-shift induced by the atom's interaction with the field in C which is characterized by the Rabi frequency $\Omega/2\pi$=49 kHz. This phase-shift is described by an operator $\Phi(\mathbf{N},\delta)$ depending upon δ and the photon number operator $\mathbf{N}=\mathbf{a}^\dagger\mathbf{a}$ (**a** and $\mathbf{a}^\dagger$: photon annihilation and creation operators). To lowest order, $\Phi(\mathbf{N},\delta)$ is linear in **N**, but for the small $\delta/\Omega$ values of our experiment, we take into account its exact non-linear expression[13]. The interferometer measures $\cos(\Phi(\mathbf{N},\delta) + \phi)$ which is sensitive to the diagonal elements of the field density matrix in the Fock state basis, but tells nothing about the coherences between these states. To get this information, we measure the phase-shifts produced by the field after it has been translated in phase space, by mixing it with reference coherent fields of adjustable complex amplitudes α. These translations, described by the operators $\mathbf{D}(\alpha) = \exp(\alpha\mathbf{a}^\dagger - \alpha^*\mathbf{a})$, are achieved by injecting a second field pulse in C.

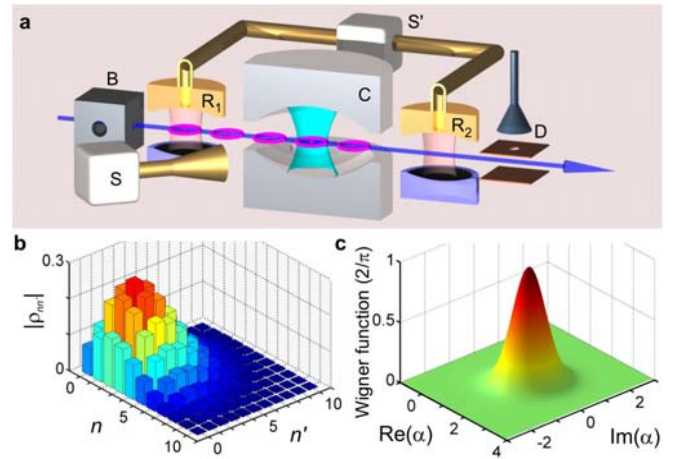

**Figure 1 | Reconstructing a coherent state.** a, Sketch of the set-up showing the stream of atoms prepared in box B and crossing the $R_1$-$R_2$ interferometer in which the cavity C, made of two mirrors facing each other, is inserted. The source S coupled to a waveguide generates a coherent microwave pulse irradiating C on the side. By diffraction on the mirrors' edges, it injects in C a small coherent field with controlled amplitude and phase. The outside field vanishes quasi-instantaneously after S is switched off. The source S is used to prepare an initial field in C and, later, to translate the field for state reconstruction. Another pulsed source S' feeds the interferometer cavities $R_1$ and $R_2$. Information is extracted from the field by state selective atomic counting in D. **b**, Density matrix (absolute values of matrix elements) of a coherent state of amplitude β = √2.5, reconstructed in an 11-dimension Hilbert space. The reconstruction parameters are $\delta/2\pi$ = 65 kHz and $\phi = -\Phi(0,\delta)+\pi$. We sample 161 points in phase space and for each point detect ≈7,000 atoms over 600 realizations. The fidelity F=$\langle\beta|\rho|\beta\rangle$ of the reconstructed state is 0.98. **c**, Wigner function (in units of 2/π) obtained from the density matrix shown in b.

[1]Laboratoire Kastler Brossel, Département de Physique de l'Ecole Normale Supérieure, CNRS and Université Pierre et Marie Curie, 24 rue Lhomond, 75231 Paris Cedex 05, France
[2]Collège de France, 11 place Marcelin Berthelot, 75231 Paris Cedex 05, France

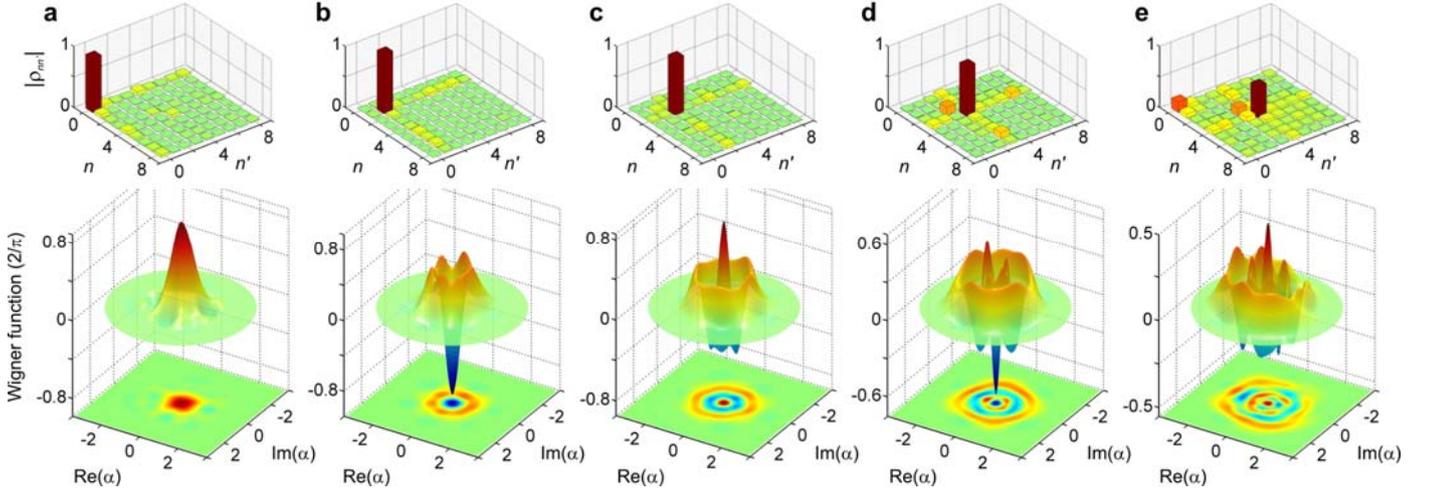

**Figure 2 | Reconstructing Fock states. a-e**, Reconstructed density matrices (absolute values of matrix elements) and Wigner functions (in units of $2/\pi$) of the $n_0=0$ to $n_0=4$ Fock states prepared by quantum non-demolition projection of an initial coherent field ($n_m=1.5$ for $n_0=0\ldots3$ and $n_m=5.5$ for $n_0=4$). The Wigner functions (in units of $2/\pi$) are shown as 3D plots and 2D projections. We select a photon number $n_0$ if, after the detection of ~60 preparation atoms, the measurement has converged to a Fock state having a probability >0.9 for $n_0=0\ldots3$, and >0.8 for $n_0=4$. The same detuning $\delta/2\pi = 120$ kHz is used for state preparation and reconstruction, corresponding to $d\Phi/dn \approx \pi/2$ at $n=3$. Two values of $\phi$ ($-\Phi(0,\delta)+\pi$ and $-\Phi(0,\delta)+\pi/2$) are used for state preparation and reconstruction, which is made in a 9-dimension Hilbert space. We sample ~400 points in phase space and, for each point, average between about 50 (for $n_0=3$) and 500 realizations ($n_0=0$ and 1) with ~10 atoms in each realization. In addition to the main peak in the density matrices, we observe for $n_0>1$ a small diagonal peak at $n=n_0-1$, due to cavity damping during reconstruction. A peak at $n=0$ also appears in the $n_0=4$ density matrix, because of imperfections in the state preparation process which selects the photon number modulo 4 (since $\Phi(n+4,\delta) \approx \Phi(n,\delta)+2\pi$). The off-diagonal elements in the density matrices and the corresponding fluctuations in the angular distributions of the WFs mainly reflect statistical noise (less atoms detected for reconstructing $n_0 = 2$, 3 and 4 than for $n_0 = 0$ and 1). The fidelities $F = \langle n_0|\rho|n_0\rangle$ of the reconstructed states are 0.89, 0.98, 0.92, 0.82, 0.51 for $n_0=0$ to 4, respectively.

We call $\rho$ the density matrix of the field to be reconstructed (matrix elements $\rho_{nn'}$), $\rho^{(\alpha)} = \mathbf{D}(\alpha)\rho\mathbf{D}(-\alpha)$ the density matrix after field translation and $P_e$ ($P_g$) the probability for finding in $|e\rangle$ ($|g\rangle$) the first atom having crossed the interferometer (experimentally obtained by averaging over many field realizations). The difference $P_e-P_g = \text{Tr}[\rho^{(\alpha)}\cos(\Phi(\mathbf{N},\delta)+\phi)]$ is the expectation value in the translated state of the diagonal field operator $\cos[\Phi(\mathbf{N},\delta)+\phi]$. The measurement is non-demolition for the photon number[15] and the ensemble average of first crossing atoms does not change $\rho^{(\alpha)}_{nn}$. Hence, the same $P_e-P_g$ expression holds for the second (or any subsequent) atom. We thus determine $P_e-P_g$ by averaging the detections of successive atoms along a single field realization together with those coming from different realizations. A measuring sequence on each realization lasts 4 ms, a time short compared to the state characteristic evolution time. We also correct the raw $P_e-P_g$ values by taking into account the known imperfections of the interferometer.

The $P_e-P_g$ difference is also the expectation value of $\mathbf{G}(\alpha,\phi,\delta) = \mathbf{D}(-\alpha)\cos[\Phi(\mathbf{N},\delta)+\phi]\mathbf{D}(\alpha)$ in state $\rho$. By sampling $\alpha$ values, we obtain the expectations $g(\alpha,\phi,\delta)$ of an ensemble of non-commuting $\mathbf{G}(\alpha,\phi,\delta)$ operators satisfying:

$$\text{Tr}[\rho \mathbf{G}(\alpha,\phi,\delta)] = g(\alpha,\phi,\delta). \quad (1)$$

Provided we sample a large enough number of $\alpha$-points in phase space, formula (1) allows us to reconstruct $\rho$. To insure that the reconstructed state does not contain any information other than that extracted from the data, we also maximize the field entropy $-\text{Tr}[\rho \ln\rho]$ during the reconstruction procedure (principle of maximum entropy[16]).

The Wigner function (WF) associated to state $\rho$ is defined at point $\alpha$ in phase space as[14]

$$W(\alpha) = 2\text{Tr}[\mathbf{D}(-\alpha) \rho \mathbf{D}(\alpha) e^{i\pi\mathbf{N}}]/\pi \quad (2)$$

and (to within a normalization) is the expectation value of the photon number parity operator $\exp(i\pi\mathbf{N})$ in the state translated by $-\alpha$. The WF could be determined directly[17] if the atoms underwent an exact phase shift of $\pi$ per photon, realizing the measurement of $\exp(i\pi\mathbf{N})$ after field translation by different $\alpha$'s. This would be a special case of reconstruction corresponding to $\Phi(\mathbf{N},\delta) - \Phi(0,\delta) = \pi\mathbf{N}$. This relation cannot be satisfied due to non-linear atom-field phase-shift. Instead of a direct determination of the WF, we thus reconstruct $\rho$ first, then obtain the WF using formula (2).

Figure 1b shows the reconstructed density matrix of a coherent state. Along its diagonal, we recognize the Poisson photon number distribution $\rho_{nn}$. The off-diagonal elements describe the classical coherence of the state. The corresponding WF, shown in Fig 1c, is, as expected, a Gaussian peak with a circular symmetry.

As a first non-classical example, we have reconstructed Fock states. To generate them, we prepare a coherent field and let it interact with atoms, achieving a quantum non-demolition measurement of the photon number which progressively projects the field onto a Fock state $|n_0\rangle$. This procedure is adapted from[15], taking into account the known effect of cavity damping during state projection. Following this preparation, we apply our reconstruction method with subsequent probe atoms and reconstruct the Fock states present in the expansion of the initial coherent state.

Figure 2 displays the obtained density matrices together with the corresponding WFs for $n_0 = 0$ (vacuum), 1, 2, 3 and 4. As expected, the density matrices mainly exhibit a single diagonal peak. Each WF shows circular rings around phase space origin, where it is positive for even $n_0$, negative for odd $n_0$. The number of rings and their size increases as expected with $n_0$. Photonic Fock states with small $n_0$ have already been reconstructed in free-space[18-20] or in a cavity[21], but this is to our knowledge the first Fock state reconstruction with $n_0 > 2$.

To generate a SC state[22], we first inject in C a coherent field of amplitude $\beta = \sqrt{n_m}$. We then prepare an atom in the state $(|e\rangle + |g\rangle)/\sqrt{2}$ using $R_1$ and send it into C. The two atomic components shift the phase of the field in opposite directions. Neglecting atom-field phase shift non-linearity, the field is split into two coherent states of complex amplitudes $\beta\exp(\pm i\chi)$, where $\chi = [d\Phi(n,\delta)/dn]/2$ evaluated at $n = n_m$. The atom is entangled with the field in the state $(|e\rangle|\beta\exp(i\chi)\rangle + |g\rangle|\beta\exp(-i\chi)\rangle)/\sqrt{2}$. The $R_2$ pulse mixes again $|e\rangle$ and $|g\rangle$. Finally the atomic detection, depending upon its outcome ($|g\rangle$ or $|e\rangle$), projects the field into one of the two SC states ($|\beta \exp(i\chi)\rangle \pm |\beta \exp(-i\chi)\rangle)/\sqrt{2}$. We call them "even" (+ sign) and "odd" (- sign) SC states since they contain, for $\chi = \pi/2$, only even and odd photon numbers, respectively. After this preparation, we apply our reconstruction procedure.

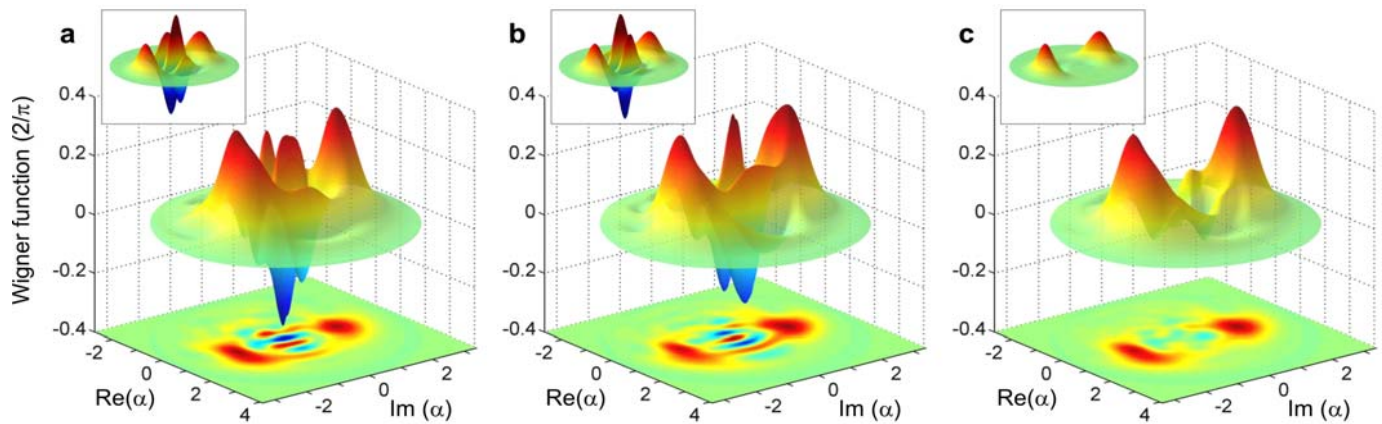

**Figure 3 | Reconstructing Schrödinger cat states.** The WFs of even (**a**) and odd (**b**) SC states (in units of $2/\pi$) with $n_m$=3.5 and $\chi$=0.37$\pi$ are reconstructed, following state preparation. The same detuning ($\delta/2\pi$ = 51 kHz) and interferometer phase ($\phi = -\Phi(0,\delta)+\pi$) are used for state preparation and reconstruction. The number of sampling points is $\approx$500, with $\approx$2,000 atoms detected at each point, in 400 realizations. The dimension of the Hilbert space used for reconstruction is 11. The small insets present for comparison the theoretical WFs computed in the case of ideal preparation and detection of the atomic state superpositions. Decoherence during state preparation is taken into account. The maximum theoretical values of the classical components and interference fringes are close to 0.5 and 1, respectively. In the reconstructed states, the quantum interference is smaller, mainly due to imperfections of the Ramsey interferometer which affect the cat state preparation (and not its reconstruction). **c**, Reconstructed WF of the field prepared in C when the state of the preparation atom is not read-out (statistical mixture of two classical fields). In the inset: corresponding theoretical WF.

Figure 3a and b shows the WFs of the even and odd cat states obtained from the same coherent field ($n_m$ = 3.5 and $\chi$ = 0.37 $\pi$). They exhibit two well-separated positive peaks associated to the classical components, whose slightly elongated shape is due to the phase-shift non-linearity neglected above. The "size" of each SC state, defined as the squared distance between peaks, is $d^2 \approx 4n_m \sin^2\chi$ = 11.8 photons. Between these peaks, oscillating features with alternating positive and negative values are the signatures of the SCs quantum interference. The even and odd SCs have nearly identical classical components and only differ by the sign of their quantum interference. The theoretical WFs, taking the SC preparation non-linearity into account, are shown for comparison in the insets. The fidelity of both cat states (overlap between the reconstructed $\rho$ and the expected one) is F = 0.72. It is mainly limited by imperfections of the $R_1$ and $R_2$ pulses applied to the preparation atom, which reduce the contrast of the quantum interference feature. If the preparing atom is detected without discriminating $|e\rangle$ and $|g\rangle$, we get the statistical mixture of even and odd SCs whose WF is shown on Fig. 3c. This is, equivalently, a statistical mixture of the two classical components. Although non-classical states of propagating light with similar WFs have been observed[23], here well-separated classical components of a field can be identified in a reconstructed state and unambiguously distinguished from their quantum interference term.

Schrödinger cats are paradigmatic states for exploring decoherence, the phenomenon accounting for the transition between quantum and classical behaviours[5]. Our reconstruction method allows us to study this process. Immediately after state preparation, we realize the $\mathbf{D}(\alpha)$ translation and detect a sequence of atoms divided into 4 ms-long time-windows. These atoms record $P_e - P_g$ versus time, without modifying the dynamics of this quantity. We average the results of realizations corresponding to the same translation and time window, and then repeat the process for different values of $\alpha$. This directly records the evolution of the translated states, rather than the one of the state itself. The two dynamics are however closely related. Decoherence acting on the initial density operator $\rho(0)$ turns it at time t into $\rho(t) = \mathbf{L}[\rho(0),t]$ where $\mathbf{L}$ is the decoherence super-operator[14] which can be shown to satisfy the relation: $\mathbf{D}(\alpha \exp[-t/2T_c])\, \mathbf{L}[\rho(0),t]\, \mathbf{D}(-\alpha \exp[-t/2T_c]) = \mathbf{L}[\mathbf{D}(\alpha)\rho(0)\mathbf{D}(-\alpha),t]$. Translating the initial field by $\alpha$ and letting it evolve during time t is equivalent to letting it evolve during that time and translating it by $\alpha \exp(-t/2T_c)$. We thus analyze the data obtained at time t as if they corresponded to a translation rescaled by $\exp(-t/2T_c)$. This is more efficient than leaving the field evolve before translating it, because we exploit all the data of a long sequence, instead of recording only a short time window for each delay. We have experimentally checked the equivalence between the two methods by comparing the results for one time delay and verified that the reconstructed SC states are, within noise, undistinguishable.

Figure 4a shows four snapshots of an odd SC WF at increasing times which clearly reveal decoherence. While the classical components have hardly decayed, the interference feature has vanished after 50 ms, turning the initial state into a statistical mixture similar to that shown in Fig. 3c. A complete movie of a SC WF evolution is presented as supplementary information. By subtracting the WF of the even and odd SCs corresponding to the same preparation sequence, we isolate their interference features by cancelling their equal, classical, parts. A movie showing the progressive vanishing of this difference is also provided as supplementary information.

It is also instructive to observe decoherence directly on the density matrix. In order to distinguish the classical coherence of each SC component from their mutual quantum coherence, we consider the mathematically translated reconstructed density matrix $\rho^T = \rho^{(-\beta \exp(i\chi))}$ whose classical components are close to the vacuum $|0\rangle$ and to $|-2i\beta\sin\chi\rangle$. This formal translation leaves unchanged the distance of the two classical components in the phase plane as well as their mutual coherence.

In Fig. 4b, we present the density matrix $\rho^T(t)$ of the SC state in Fig. 4a, reconstructed for the same times. In each frame, the diagonal elements present two maxima around n = 0 and n = 11. The off-diagonal elements are of two kinds. Those for which $|n-n'| \approx \sqrt{11}$ describe the classical coherence of the non-vacuum component and remain nearly unchanged on the observed timescale. The off-diagonal terms in the first row and column of the matrix (respectively $\rho^T_{0n}$ and $\rho^T_{n0}$) initially exhibit a bell-shaped variation with n, centred at n $\sim$11. These terms correspond to the SC quantum coherence responsible for the oscillations observed in the WF. Their fast decay is the signature of decoherence.

The measured quantum coherence of the even and odd cats is plotted versus time in Fig. 4c. A common exponential fit yields a decoherence time $T_d = 17 \pm 3$ ms. A simple analytical model of decoherence[14] predicts $T_d = 2T_c/d^2 = 22$ ms at T = 0 K, reduced to[24] $T_d = 2T_c/(d^2(1+2n_b)+4n_b) = 19.5$ ms when including thermal background at T = 0.8 K, in good agreement with the measured value. A movie of a smaller SC ($d^2 = 8$) yields $T_d = 28$ ms, illustrating the dependence of the decoherence time on the cat size[5,14]. Earlier experiments have studied the relaxation of photonic[22] and atomic[25] SCs by observing specific features of their states, but this experiment is the first to realize a movie of decoherence on a fully reconstructed SC.

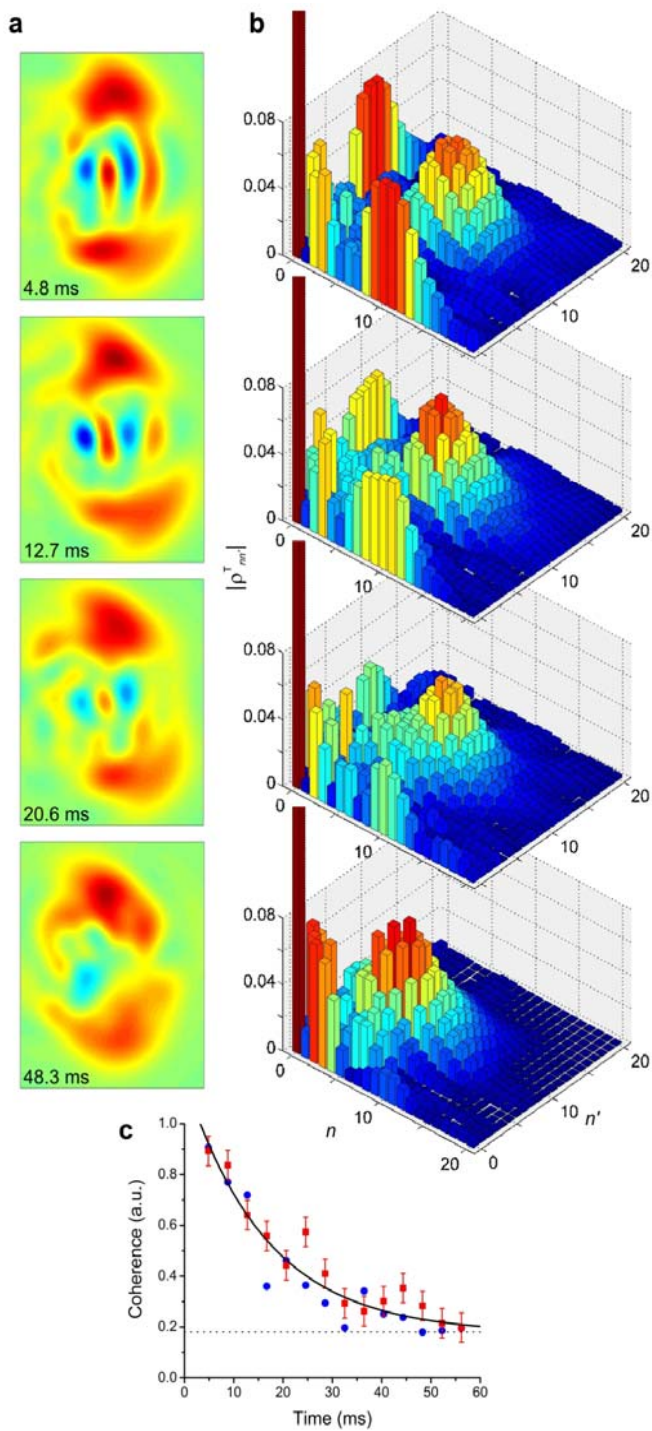

**Figure 4 | Movie of decoherence. a**, Snapshots of the odd cat WF of Fig. 3b at four successive times after state preparation. **b**, Corresponding snapshots of the translated density matrix $\rho^T$ (absolute value of matrix elements). The size of the Hilbert space is enlarged to a dimension of 30. The n=0 peak is clipped, its amplitude being ≈ 0.38 in each snapshot. In all frames, the small matrix elements for n<5 are due to the deviation of the SC classical component from an ideal coherent state. **c**, The quantum coherence of the SC states, defined as the sum of the $\rho^T_{n0}$ for n ≥ 5, is shown versus time for the even (red squares) and odd (blue circles) SCs. The statistical error bar (shown for clarity for the even SC state only) is obtained by analysing the state reconstructions performed on different sub-samples of measured data. The solid line is a common exponential fit, including an offset (dotted line) accounting for residual noise in the absolute values of the density matrix elements.

We have shown that atoms interacting with a cavity field can be used to engineer and reconstruct a wide variety of photonic states and to study their evolution. Pushing one step further, we plan to use information provided by the atoms to implement feedback procedures and preserve the quantum coherence over longer time intervals[26]. We will also extend these studies to fields stored in two cavities. Atoms will be used to entangle the cavity fields into non-local quantum states[27,28], reconstruct these states and protect them against decoherence by quantum feedback operations.

**Acknowledgements** This work was supported by the Agence Nationale pour la Recherche (ANR), by the Japan Science and Technology Agency (JST), and by the EU under the IP projects SCALA and CONQUEST. S.D. is funded by the Delegation Generale à l'Armement (DGA).
**Author Contributions** S.D., I.D. and C.S. contributed equally to this work.
**Author Information** Correspondence should be addressed to S.H. (haroche@lkb.ens.fr).